\documentclass[aps,prl,amsmath,amssymb,a4paper,twocolumn,showpacs]{revtex4-1}

\usepackage{amsmath}    
\usepackage{graphicx}   
\usepackage{verbatim}   
\usepackage{color}      
\usepackage{hyperref}   
\usepackage{nicefrac}
\usepackage{lipsum}
\usepackage{upgreek}
\usepackage{notes2bib}
\usepackage{natbib}

\usepackage[normalem]{ulem}

\newcommand{\p}{{\boldsymbol p}}

\newcommand{\C}{{\boldsymbol C}}

\newcommand{\F}{{\boldsymbol F}}
\renewcommand{\S}{{\boldsymbol S}}

\begin{document}

\title{Optimal Transport Flows for Distributed Production Networks}
\author{Julius B. Kirkegaard}
\author{Kim Sneppen}

\affiliation{Niels Bohr Institute, University of Copenhagen, 2100 Copenhagen, Denmark}

\date{\today}

\begin{abstract}
Network flows often exhibit a hierarchical tree-like structure
that can be attributed to the minimisation of dissipation.
The common feature of such systems is a single source 
and multiple sinks (or vice versa).
In contrast, here we study networks with only
a single source and sink.
These systems can arise from
secondary purposes of the networks,
such as blood sugar regulation through insulin production.
Minimisation of dissipation in these systems
lead to trivial behaviour.
We show instead how optimising the transport time
yields network topologies that
match those observed in the insulin-producing pancreatic islets.
These are patterns of periphery-to-center
and center-to-periphery flows.
The obtained flow networks are broadly independent of
how the flow velocity depends on the flow flux,
but continuous and discontinuous phase transitions appear
at extreme flux dependencies.
Lastly, we show how constraints on flows
can lead to buckling of the branches of the network,
a feature that is also observed in pancreatic islets.
\end{abstract}

\maketitle

Transport networks are essential for life to function
on large multicellular scales.
In vertebrates, blood flow delivers energy and nutrients, and removes waste
through the branched network of the vascular system.
The separation of vessels into arteries and veins
ensures that oxygen-rich and oxygen-depleted parts of the network
are kept separate.
In plants the separation into xylem and phloem provides a similar function.
In nature, these systems typically exhibit tree-like hierarchical structures,
as \textit{e.g.} in the aorta which
splits into increasingly smaller arteries all the way to
capillaries, the smallest vessels of the circulatory system.
The tree structure can be understood as
minimising the dissipation of the blood flow through
the system \cite{Bohn2007, Hu2013, Ronellenfitsch2016}.
Furthermore, loop-redundant tree-like structures,
as is evident from observing \textit{e.g.} the veins of a tree leaf,
is explained by minimising dissipation 
under damage or under fluctuating needs \cite{Katifori2010, Corson2010}.
Perhaps because of its origin as a power-minimising network,
tree-like structures are observed not just in vascular networks,
but also for instance
in both natural and artificial river networks \cite{Dodds2010, Katifori2010}.
The shared feature of these systems is that
they consists of a single fluid source with a lot of sinks,
or a single sink with a lot of sources
(these are equivalent, dual formulations).
For instance, in the arterial system the heart is the source
and the body cells the sinks,
whereas the roles are reversed in venous systems.

The study of flow networks that have a single source and a single sink
have received much less attention,
perhaps because it is not clear what properties can be optimised over such networks
--- indeed dissipation minimisation lead to trivial behaviour.
However, such networks are critical in flow systems that have secondary products
(\textit{e.g.} other than oxygen in blood) that are being produced or delivered
along the flow path.
Here, we exemplify such a system by the Islets of Langerhans
in the pancreas.
In these islets, beta cells release insulin and alpha cells glucagon
into the blood stream based on blood glucose levels \cite{Pour2002, Hong2013}.
In such systems there is no need for an arterial-venous separation,
and the transport from any hormone-producing cell
directly couples to both the downstream
and upstream transport systems.

\begin{figure}[b]
    \centering
    \includegraphics{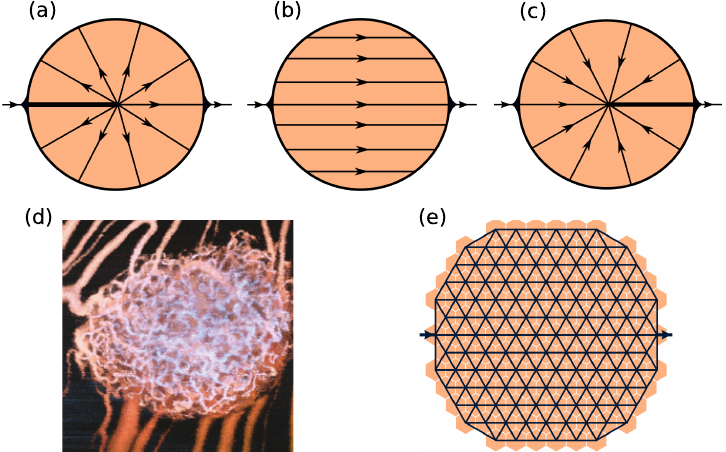}
    \caption{Vasculature of pancreatic islets.
    (a) Center-to-periphery flow.
    (b) Left-right symmetric flow.
    (c) Periphery-to-center flow.    
	(d) Microscopy of vasculature, from Ref. \cite{Berclaz2016},
	showing the high degree of tortuousness
	in the vasculature of pancreatic islets.
	(e) The network structure considered.
    This is constructed by taking a circular subsection
    of a hexagonal grid of nodes. Neighbours are then
    induced from a Delaunay triangulation. 
    Each edge has a conductivity $C_e$ associated to it,
    which are the parameters to be optimised over.
    Inlet and outlet are shows by arrows.}
    \label{fig:graph}
\end{figure}

The vasculature of pancreatic islets differs
from species to species.
In particular, various topologies have been observed:
periphery-to-center flow, straight through, and center-to-periphery flow,
as illustrated in Fig. \ref{fig:graph}(a-c) \cite{Shimokawa1988, Nyman2008, El-gohary2018}.
For instance in rodents, the center-to-periphery topology is the most common \cite{Nyman2008}.
furthermore, the vasculature of these islets is often very tortuous
compared to the vasculature of other organs \cite{Berclaz2016, Kirkegaard2019}
as shown in Fig. \ref{fig:graph}d.
In this Letter we propose an optimisation problem over single-source, single-sink flows
and use this to provide possible explanations
for the type of features observed in pancreatic islets.

\begin{figure}[b]
    \centering
    \includegraphics{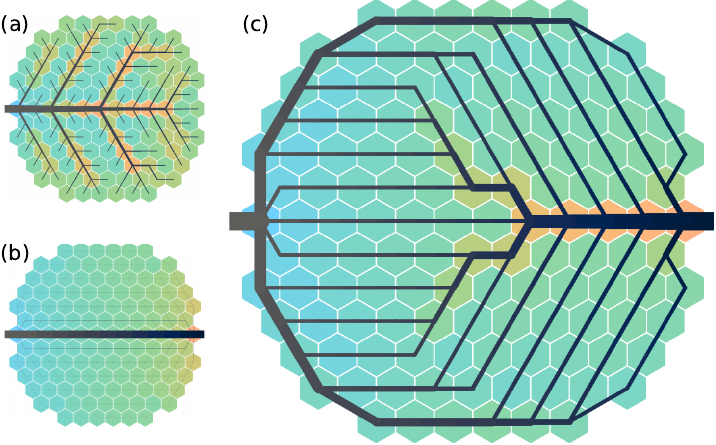}
    \caption{Optimised network structures.
    (a) Minimal dissipation network with sinks at all nodes.
    (b) Minimal dissipation network with a single sink at edge.
    (c) Per-node time minising network with a single sink at edge.
    Cell colours indicate product concentration flowing through that node
    as calculated by solving $\rho_i = 1 + \sum_{j \in \mathcal{I}_i} |F_{ij}| \, \rho_j / \sum_{k \in \mathcal{O}_j} |F_{jk}|$
    ($\mathcal{I}_i$ denoting incoming and $\mathcal{O}_i$ outgoing nodes of node $i$).    
    Flow lines are coloured by pressure,
    their thickness indicating (square root of) flow magnitude.}
    \label{fig:optimum}
\end{figure}

\textit{Model} --- 
To study systems of blood flow optimisation we consider,
as in previous studies \cite{Bohn2007, Hu2013, Katifori2010, Ronellenfitsch2016, Corson2010},
flows on networks.
While pancreatic islets indeed can have more than one inlet and outlet,
we simplify the system and consider the network shown in Fig. \ref{fig:graph}e.
Nevertheless, our approach works for any number of sources and sinks.
In Fig. \ref{fig:graph}e cells are represented by hexagons,
and edges betweens between these cells indicate where fluid can potentially flow.
Inlet and outlet are indicated by arrows.

This specific graph has $n_N = 130$ nodes and $n_E = 357$ edges,
and each edge $e$ of the graph has a conductivity $C_e$ and a length $L_e$.
An $n_E \times n_N$ oriented incident matrix $\Delta$ of the graph
gives each edge a unique direction,
and we can thus attach to each edge a (signed) flow $F_e$.
We furthermore define the source vector $\S$ 
with $S_\text{source} = 1$, $S_\text{sink} = -1$
and $S_i = 0$ elsewhere
and require that the flow obeys Kirchoff's current law
\begin{equation}
\Delta^T \F = \S.
\label{eq:kirchoff}
\end{equation}
Since $n_N < n_E$, the flow is far from determined
by this condition alone.

To make the flow well-defined,
we furthermore require that it is derivable from a potential
based on the effective conductivities,
\begin{equation}
\F = \C^{\text{eff}} \Delta \p,
\label{eq:potential}
\end{equation}
where $p_i$ is the potential (pressure) defined at the nodes,
and $\C^{\text{eff}}$ is a diagonal matrix
with entries $C^{\text{eff}}_{ee} = C_e / L_e$.
Combining Eq. \eqref{eq:kirchoff} and \eqref{eq:potential}
we can solve for the potential as
\begin{equation}
\p = [ \Delta^T \C^{\text{eff}} \Delta ]^\dagger \, \S,
\end{equation}
where $\dagger$ denotes the pseudo-inverse,
which is needed since the system of equations
is singular, but can be solved by the pseudo-inverse if $\sum_i S_i = 0$,
which is indeed the case here.
From the potentials $\p$, the flows are immediately obtained
from Eq. \eqref{eq:potential}.

The total power dissipation of the system is
$P = \sum_e F_e^2 / C^{\text{eff}}_{ee}$ \cite{Bohn2007, Hu2013, Ronellenfitsch2016}.
As mentioned, minimising this term leads to tree topologies
for a single source and sinks everywhere
($S_\text{source} = 1$, $S_i = -1/(n_N - 1)$ elsewhere).
This optimum on our network topology
is shown in Fig. \ref{fig:optimum}a.
The minimisation is done under constant ``material cost'' $\sum_e L_e C_e^\gamma$,
and the tree-like structures are obtained for $\gamma < 1$.
In the optimum, the conductivities scale with the flow as
\bibnote{Briefly, minimising $\sum_e L_e F_e^2/C_e$ under constant $\sum_e L_e C_e^\gamma$ by the method of Lagrange multipliers leads to $-F_e^2/C_e^2 + \lambda \gamma C_e^{\gamma - 1} = 0$ from which the scaling follows. See \textit{e.g.} Ref. \cite{Bohn2007} for details.}
\begin{equation}
C_e \sim |F_e|^{2 / (1 + \gamma)},
\label{eq:powerscaling}
\end{equation}
\textit{i.e.} a larger conductivity is needed where there is a lot of flow.

Trivially, minimising power dissipation
in the network with just a single
source and single sink leads to flow only going along
the shortest path between the source and sink
as shown in Fig. \ref{fig:optimum}b.
This is also, naturally, the time minimising network for
flow between the source and the sink.
We are interested in network structures that visit all
nodes in an ``optimal'' way.
Indeed proper distribution of vessels is far more important in
pancreatic islets and other systems
than the (potentially miniscule) power being dissipated.

We consider instead graphs that minimise the average time for
the product (\textit{e.g.} insulin) being produced at the nodes
to reach the outlet
(the opposite inlet-centric definition will be discussed later).
This time-optimised graph we define as follows:
Denote for each node $T_i$ the average time taken from that node
for product to reach the outlet.
The average time that we intend to minimise is thus
\begin{equation}
\langle T \rangle = \frac{1}{n_N} \sum T_i,
\label{eq:taverage}
\end{equation}
where $T_i$ is found by the recursive relation
which follows by letting product flow in proportion
to the fluid flow,
\begin{equation}
T_i = \frac{ \sum_{j \in \mathcal{O}_i} | F_{ij} | \, (T_j + T_{ij})}{\sum_{j \in \mathcal{O}_i} | F_{ij} |},
\label{eq:ti}
\end{equation}
with the special case $T_\text{sink} = 0$.
This is a linear equation for $T_i$,
where $\mathcal{O}_i$ is the set of nodes
that are outgoing from node $i$.
Whether an edge $e$ is outgoing from node $i$ can be identified by
the criteria $F_e \, \Delta_{ei} > 0$.
$T_{ij}$ is the time taken for the product to flow
from node $i$ to neighbouring node $j$.
The physics of the system is determined through the relation between $T_{ij}$
and $F$, $C$ and $L$.

\begin{figure}[b]
    \centering
    \includegraphics{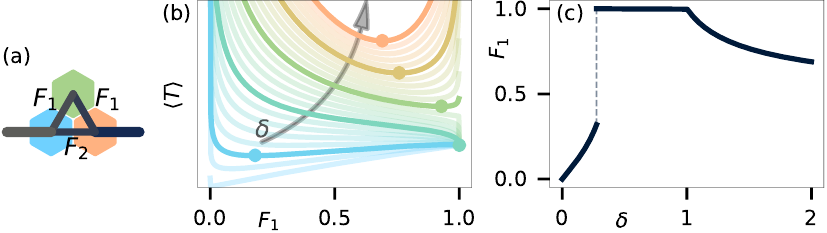}
    \caption{Phase transition in triangle geometry.
    (a) Triangle geometry. We consider
    $S_\text{source} = 1 = F_1 + F_2$.
    (b) Value of $F_1$ obtained by minimising $\langle T \rangle$
    for various scalings $\delta$.
    There is a discontinuous transition at $\delta \approx 0.275$
    and a continuous transition at $\delta = 1$.
    (c) Energy ($\langle T \rangle$) landscape
    for various $\delta$.
    Circles indicate select minima.}
    \label{fig:phasetrans}
\end{figure}

For blood vessels each edge corresponds to a tube of a given radius $r_e$.
We can thus relate $F_e \sim v_e r_e^2$,
where $v_e$ is the fluid velocity along the tube.
Furthermore, in Poiseuille flow $C_e \sim r_e^4$,
and thus $T_{ij} = T_e = L_e/v_e \sim L_e \sqrt{C_e} / F_e$.
This shows that the time travelling across an edge is minimised
by having a small conductivity, since in small tubes the liquid
will have to move faster for the same flux $F$.
So while we are interested in solutions that minimise time,
this formulation yields solutions that are severely
dissipation inefficient
in which case power optimisation becomes more relevant than time minimisation.
Instead, we reformulate $T_{ij}$ in terms of only the flows $F$
such that we stay in reasonably power-efficient regimes,
as established by Eq. \eqref{eq:powerscaling}.
Taking $\gamma = 1$, we have
\begin{equation}
T_{ij} \sim \frac{L_{ij}}{\sqrt{|F_{ij}|}},
\label{eq:tij}
\end{equation}
which is the definition we will use,
and critically what enables the optimisation problem
to yield non-trivial results.
Variations induced by different values of $\gamma$ also work,
as will be shown.
We thus minimise the average time $\langle T \rangle$
in the regime where large fluid flows are associated
with large conductivities.
Equations (\ref{eq:taverage}--\ref{eq:tij}) define the optimisation problem,
which we solve by a momentum-based version of gradient descent \cite{Note2}.

\textit{Periphery-center optima} --- 
The result of our minimisation scheme is shown in Fig. \ref{fig:optimum}c.
First we note that the single-source-sink system prevents self-similar
branching solutions that are known from single-source, multiple-sink systems.
Instead the solution only has a few main branches, in particular
one at the periphery and one at the center.
Interestingly, this exact pattern is one of the three
topologies [Fig. \ref{fig:graph}(a-c)] of pancreatic islet blood flow observed in nature
\cite{Nyman2008, El-gohary2018}.

In some species, such as rodents,
the glucagon-producing alpha cells
and the insulin-producing beta cells comprising
the islets are heterogeneously distributed with the beta cells in the center
and alpha cells at the periphery.
It has thus been suggested that the order of the flow
suggests intercellular communication and regulation \cite{Nyman2008},
\textit{i.e.} beta cells regulating alpha cells or vice versa,
depending on the flow topology.
Our results show a separate, but non-exclusive explanation
that the patterns can appear due to an optimisation of the flow itself,
independent of any heterogeneous distribution of cells.
Interpreted in an evolutionary sense,
one could imagine that optima such as the one
derived here led to the present flow topology
before the advent of alpha and beta cells
and thus drove the heterogeneous distribution thereof.

The equal distribution of flow to each node
and the sparsity of the network in the optimum [Fig. \ref{fig:optimum}c]
is possible due to the scaling of Eq. \eqref{eq:tij}.
In fact, considering instead $T_{ij} \sim L_{ij} / |F_{ij}|^\delta$
our model works for a broad range of $\delta$.
To illustrate this, consider the simple network of three nodes shown in
Fig. \ref{fig:phasetrans}a.
Taking $1 = F_1 + F_2$ it follows that $2 \langle T \rangle = (1 - F_1)^{1-\delta} + 2 F_1^{1 - \delta} + F_1^{-\delta}$,
which is shown in Fig. \ref{fig:phasetrans}b for various values of $\delta$.
Minimising $\langle T \rangle$, the optimal $F_1$
is shown in Fig. \ref{fig:phasetrans}c,
which demonstrates that this system has a discontinuous phase transition
at $\delta \approx 0.275$ and
a continuous phase transition at $\delta = 1$ in $F_1$.
Between these values the optimal $F_1$ is independent of $\delta$,
which is the regime we study.

The discontinuous phase transition occurs because
the left and top node of Fig. \ref{fig:phasetrans}a
have conflicting optima:
the left node minimises its time by having $F_2$ large,
whereas the top node needs $F_1$ large.
Large $\delta$ favour large $F_1$ because then a larger flow velocity compensates 
for the larger length of the upper branch. In contrast, 
a small $\delta$ favours the shorter path of the lower branch, 
while only leaving a small flux through the upper branch 
required to transport the product from the top node.

For more complex graphs such as the one we consider [Fig. \ref{fig:graph}e],
the phase transition behaviour is much more rich and complex
and the discontinuous transition is pushed to lower values,
and the continuous transition happens
at a $\delta$ smaller than (but close to) 1.
Our choice of $\delta = \nicefrac{1}{2}$ lies safely within
the regime, where the results
are independent of the precise value of $\delta$
and where the resulting networks are sparse graphs.
Thus our results are to a large degree independent
of the scaling in Eq. \eqref{eq:tij}.

The left-right asymmetry and
thus the periphery-to-center flow of Fig. \ref{fig:optimum}c
comes from the fact that we are minimising
the time for the product to reach the outlet from the nodes.
The large collection branch emerging from the center
thus minimises the time for many nodes
by providing a fast route.
Had we instead minimised the time from the source to the nodes $\langle T_r \rangle$
the solution would be reversed,
since it would be important to reach the nodes fast.
Indeed this opposite situation, with flow from center to periphery,
is the most commonly observed topology in rodents \cite{Nyman2008},
and could perhaps hint that the time for ``information''
(\textit{e.g.} glucose levels) to reach the cells
is more important than the product to exit the islet.

\begin{figure}[t]
    \centering
    \includegraphics{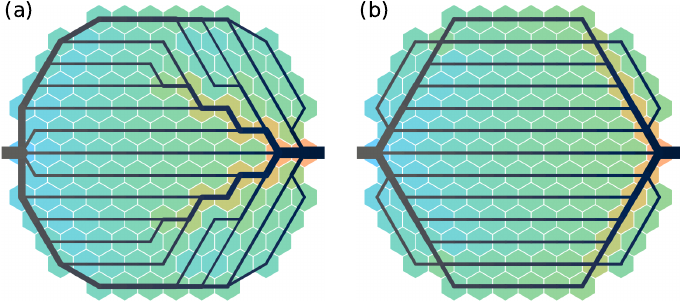}
    \caption{Varying the ratio $\alpha$ between time
    from nodes to outlet and inlet to nodes.
    (a) Obtained optimum at $\alpha = 0.25$.
    (b) The left-right symmetric solution obtained at $\alpha = 0.5$.}
    \label{fig:mix}
\end{figure}

Naturally, the combination of time from inlet to nodes and from nodes to outlet
$\langle T_c \rangle = (1 - \alpha ) \, \langle T \rangle + \alpha \, \langle T_r \rangle$ ($0 \leq \alpha \leq 1$) can also be optimised over.
Fig. \ref{fig:mix} shows examples of
minimising $\langle T_c \rangle$.
$\alpha = 0$ corresponds
to situation we have already studied,
and $\alpha = 1$ simply left-right mirrors this solution.
In-between a compromise of the two optima is reached,
\textit{e.g.} as shown for $\alpha = 0.25$ in Fig. \ref{fig:mix}a.
The value of $\langle T_c \rangle$ is
minimised for $\alpha = 0$ and $\alpha=1$,
since for any other value the solution will
be the average of two solutions
that both compromise.
At precisely $\alpha = 0.5$ the solution becomes left-right symmetric
[Fig. \ref{fig:mix}b] yielding an islet topology similar
to that of Fig. \ref{fig:graph}b.

\textit{Minimal flow constraints} ---
While the solution of Fig. \ref{fig:optimum}c does indeed
distribute the flow to all nodes, some nodes see much more fluid
flowing through them than others.
As the pancreatic islets grow 
and if the blood flow source cannot keep up with this growth,
each cell/node will see a decrease in flow permeating them.
In this way, growth in networks flows can be emulated
by a decreasing $S_\text{source}$ \cite{Ronellenfitsch2016}.
At the same time a minimum amount of flow might be required
at each cell
\textit{e.g.} for accurate estimation of glucose levels.
This leads us to consider adding
minimal flow constraints to the optimisation problem.

\begin{figure*}
    \centering
    \includegraphics{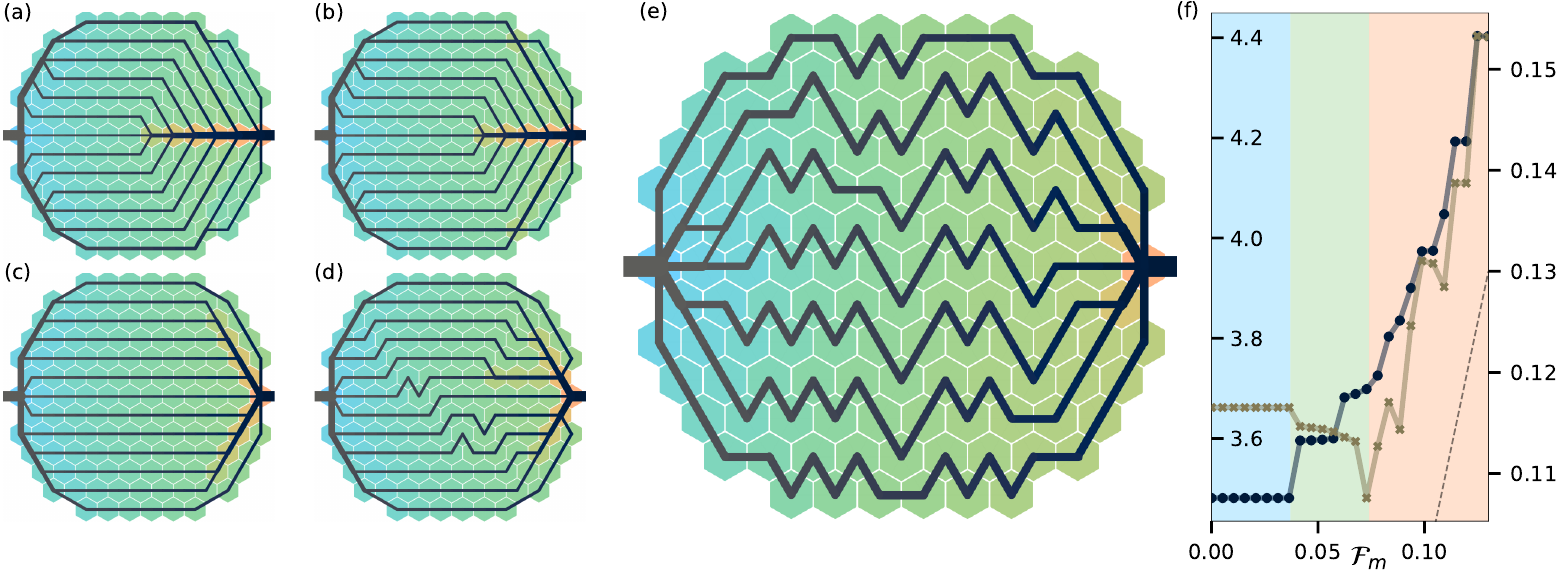}
    \caption{Optimal graphs under minimal flow constraints. 
			(a) At $\mathcal{F}_m = 0.042$, the network of Fig. \ref{fig:optimum}c loses its front collection channels.
			(b) The collection point starts moving to the right as $\mathcal{F}_m$ is increased to $0.062$.
			(c) At $\mathcal{F}_m = 0.073$ the collection point has moved to the far right.
            (d) One horizontal vessel has been removed by introducing kinks at $\mathcal{F} = 0.078$ which increases the flow in the remaining vessels.
            (e) Buckling increases; $\mathcal{F}_m = 0.125$. 
            As $\mathcal{F}_m \rightarrow 1$ the global optimum becomes a Hamiltonian path through the nodes.
            (f) In dark blue (left axis) the average time $\langle T \rangle$ is shown as function of $\mathcal{F}_m$. The average flow through the nodes $\langle \mathcal{F} \rangle$ is shown in light brown (right axis).
            Dashed line shows $f(x) = x$, which matches the slope of $\langle \mathcal{F} \rangle$ in the buckling section.
            Background colours denote section: global optimum in blue, unbuckled solution in green, and buckled in red.
            }
    \label{fig:buckle}
\end{figure*}

The flow through a node $i$ with $S_i = 0$ is
\begin{equation}
\mathcal{F} = \sum_{j \in \mathcal{O}_i} | F_{ij} | = \sum_{j \in \mathcal{I}_i} | F_{ij} |
\label{eq:fi}
\end{equation}
and we now require
$\mathcal{F} \geq \mathcal{F}_m$ for all nodes
for a given value of $\mathcal{F}_m$.
Obtained minima are shown in Fig. \ref{fig:buckle}
under variations of $\mathcal{F}_m$.

The optimal configuration of Fig. \ref{fig:optimum}c
has $\min_i \mathcal{F} = 0.041$
and thus for any $\mathcal{F}_m$ smaller than this the same solution is obtained
(blue section in Fig. \ref{fig:buckle}f).
As $\mathcal{F}_m$ is increased above this, the network adapts
to more equally divide the flow.
First the two middle branches are lost [Fig. \ref{fig:buckle}a]
and then the collection point starts moving towards the outlet [Fig. \ref{fig:buckle}b].
In the end of this process (green section in Fig. \ref{fig:buckle}f),
the collection branch has moved all the way to the right [Fig. \ref{fig:buckle}c].
As can be seen in Fig. \ref{fig:buckle}f, naturally,
the average time $\langle T \rangle$
increases as $\mathcal{F}_m$ increases.
Fig. \ref{fig:buckle}f also shows that $\langle \mathcal{F} \rangle_i$ decreases
rapidly.
So as this reordering occurs, the minimum flow rate increases at the expense of the average flow rate.

As $\mathcal{F}_m$ is increased further than
reordering can accomodate for,
buckling occurs [Fig. \ref{fig:buckle}d],
the degree of which increases as $\mathcal{F}_m$
is increased [Fig. \ref{fig:buckle}e].
As soon as buckling occurs,
$\langle \mathcal{F} \rangle$ starts increasing
as seen in Fig. \ref{fig:buckle}f.
This increase scales linearly with $\mathcal{F}_m$,
\textit{i.e.} the average flow rate, after buckling,
stays at a fixed level above $\mathcal{F}_m$.
We note that the noise in $\langle \mathcal{F} \rangle$ in Fig. \ref{fig:buckle}f
most likely indicates that our optimisation scheme in some cases
fails to find the true global optimum
but ends in neighbouring local minima.

As shown in Fig. \ref{fig:graph}d, pancreatic islets do
indeed have a severely tortuous vasculature \cite{Berclaz2016}
similar to that obtained in Fig. \ref{fig:buckle}.
The mechanical reason for buckling in real islets
could be due to growth-induced buckling \cite{Kirkegaard2019}.
Our analysis shows how such buckling
could actually be of benefit to the system under growth.

In comparing our result with blood vessels in pancreatic islets,
we implicitly assume that transport time is of main concern.
In fact the average blood flow velocity in Langerhans islets is 
$\sim 1.4$ mm/s \cite{Berclaz2016},
implying that transport across an islet takes about $\sim 0.5$ second. 
Activity of both alpha and beta cells are pulsatile,
and in-vitro experiments show coherent oscillations of whole islets 
with periods down to three seconds \cite{Valdeolmillos1989}.
In order for the organism to utilize coherent release
of insulin downstream of islets, it is
therefore plausible that time optimization
on the sub-second scale is functional. 

In conclusion, by simply varying how much
time to nodes and time from nodes is important through the
parameter $\alpha$,
our proposed optimisation problem
on network flows directly leads to flow topologies
similar to those observed in real pancreatic islet and
illustrated in Fig. \ref{fig:graph}(a-c).
We have furthermore shown how minimum flow constraints
leads to buckling and thus tortuous vessels as
found in real islets.
Our approach applies generally to all laminar network flows
that have single-source and single-sink characteristics.

\begin{acknowledgements}
This project has received funding from the European Research Council (ERC)
under the European Union's Horizon 2020 Research and Innovation Programme,
Grant Agreement No. 740704 and the Danish National Research Foundation, Grant No. DNRF116.
\end{acknowledgements}

\end{document}